\documentclass{webofc}
\usepackage[varg]{txfonts}   
\usepackage{graphics}
\usepackage{epstopdf}
\begin{document}
\title{Jet tomography in high-energy nuclear collisions}
%
%
\author{\firstname{Ben-Wei} \lastname{Zhang}\inst{1}\fnsep\thanks{Speaker, \email{bwzhang@mail.ccnu.edu.cn}} \and
        \firstname{Guo-Yang} \lastname{Ma}\inst{1}
        \and
        \firstname{Wei} \lastname{Dai}%
        \inst{2} \and
        \firstname{Sa} \lastname{Wang}%
        \inst{1}
        \and
        \firstname{Shan-Liang} \lastname{Zhang}%
        \inst{1}
}
\institute{Key Laboratory of Quark Lepton Physics (MOE) and Institute of Particle Physics,
Central China Normal University, Wuhan 430079, China
\and School of Mathematics and Physics, China University of Geosciences, Wuhan 430074, China
          }

\abstract{%
  When an energetic parton traversing the QCD medium, it may suffer multiple scatterings and lose energy.  This jet quenching phenomenon may lead to the suppression of leading hadron productions as well as medium modifications of full jet observables in heavy-ion collisions. In this talk we discuss the nuclear modification factors and yield ratios of identified meson such as $\eta$, $\rho^0$, $\phi$,  $\omega$, and $K^0_{\rm S}$ as well as $\pi$ meson at large $p_T$ in A+A collisions at the next-to-leading order (NLO) with high-twist approach of parton energy loss. Then we discuss a newly developed formalism of combing NLO matrix elements and parton shower (PS) for initial hard production with parton energy loss in the QGP, and its application in investigating massive gauge boson ($Z^0$/$W^{\pm}$) tagged jet productions and $b\bar{b}$ dijet correlations in Pb+Pb at the LHC.
}
\maketitle
\section{Introduction}
\label{intro}
The quark-gluon plasma (QGP),  a deconfined matter composed of quarks and gluons, is expected to be created in relativistic heavy-ion collisions (HIC). To study the formation and properties of the QGP,  many signatures have been proposed and studied.  From the SPS to the RHIC, and to the LHC,  hard processes become more abundant in heavy-ion collisions with the dramatically increasing center-of-mass colliding energies.  Thus it is understandable that jet quenching, as one of the most important hard probes, is attracting more and more attentions both in theory and in experiment~\cite{Gyulassy:2003mc, Qin:2015srf, Connors:2017ptx}.  Jet quenching tells that when a fast parton passing through the hot/dense QCD medium it may suffer multiple scattering with other partons in the medium and lose a lot of energy, which may lead to not only the attenuation of leading hadrons~\cite{Gyulassy:2003mc, Qin:2015srf, Liu:2006sf, Chen:2008vha,Chen:2010te, Chen:2011vt}, but also medium modifications of jet productions and jet substructures~\cite{Connors:2017ptx, Vitev:2008rz, Vitev:2009rd, He:2011pd, Dai:2012am}. In this talk I review the recent progresses made by our group of jet quenching effect on leading hadron yields
~\cite{Dai:2015dxa, Dai:2017tuy, Dai:2017piq, Ma:2018} as well as full jet observables
~\cite{Zhang:2018urd, Dai:2018mhw, Zhang:2018} in high-energy nuclear collisions.

\section{Leading Hadron Productions in HIC}
\label{Leading Hadron Productions}
When an energetic parton propagating in the QCD medium, it may lose energy due to jet-medium interactions, which may give rise to additional term in the DGLAP evolution equations and lead to an effectively medium-modified fragmentation functions (FFs).   The medium-modified FFs can be calculated in the higher twist approach as
~\cite{Wang:2001ifa, Zhang:2003yn, Zhang:2003wk, Zhang:2004qm}:
\begin{eqnarray}
\tilde{D}_{q}^{h}(z_h,Q^2) =
D_{q}^{h}(z_h,Q^2)&+&\frac{\alpha_s(Q^2)}{2\pi}
\int_0^{Q^2}\frac{d\ell_T^2}{\ell_T^2} \int_{z_h}^{1}\frac{dz}{z} \left[ \Delta\gamma_{q\rightarrow qg}(z,x,x_L,\ell_T^2)D_{q}^h(\frac{z_h}{z},Q^2)\right.
\nonumber\\
&&\hspace{+0.5 in}+ \left. \Delta\gamma_{q\rightarrow
gq}(z,x,x_L,\ell_T^2)D_{g}^h(\frac{z_h}{z},Q^2) \right] .
\label{eq:mo-fragment}
\end{eqnarray}

\begin{figure*}
\centering
\includegraphics[width=2.5in,height=3.in,angle=0]{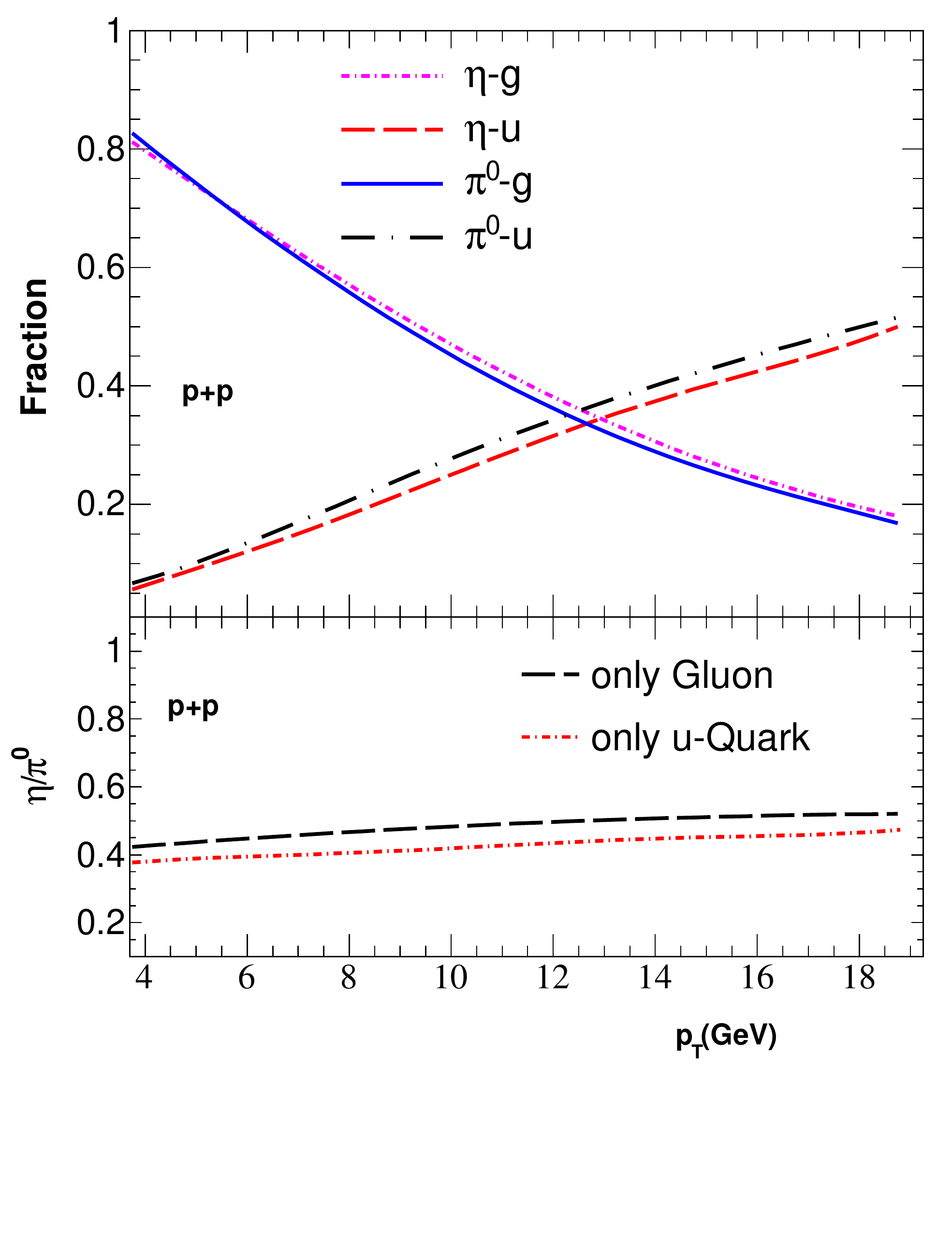}
\vspace*{-0.8cm}       
\caption{Contribution fractions of $\pi^{0}$ and $\eta$ productions from quark and gluon fragmentations in p+p collisions  at NLO with $\sqrt{s_{NN}}=200$~GeV~\cite{Dai:2015dxa}.}
\label{fig:eta-1}       
\end{figure*}

Taking into account the initial production position of fast partons and their propagation direction in the time-evolving QGP , we may obtain the averaged medium-modified FFs $\langle \tilde{D}_{c}^{h}(z_{h},Q^2,E,b)\rangle$, and cross section of the leading hadron at large transverse momentum $p_T$ in HIC at the next-to-leading order (NLO) can be given by
~\cite{Chen:2010te, Chen:2011vt, Dai:2015dxa,Dai:2017tuy, Dai:2017piq, Ma:2018}:

\begin{eqnarray}
\frac{1}{\langle N_{\rm coll}^{\rm AB}(b)\rangle}\frac{d\sigma_{AB}^h}{dyd^2p_T} &=&\sum_{abcd}\int
dx_adx_b f_{a/A}(x_a,\mu^2)f_{b/B}(x_b,\mu^2) \nonumber \\
&&\hspace{-0.5in}\times \frac{d\sigma}{d\hat{t}}(ab\rightarrow
cd)\frac{\langle \tilde{D}_{c}^{h}(z_{h},Q^2,E,b)\rangle}{\pi z_{c}}+\mathcal {O}(\alpha_s^3). \nonumber \\
\label{eq:AA}
\end{eqnarray}
In the following calculations,  EPPS16 NLO nuclear parton distribution functions (PDFs) are utilized to include initial-state cold nuclear matter effects~\cite{Eskola:2016oht}.

\begin{figure}[htbp]
\begin{center}
\includegraphics[width=2.5in,height=2.in,angle=0]{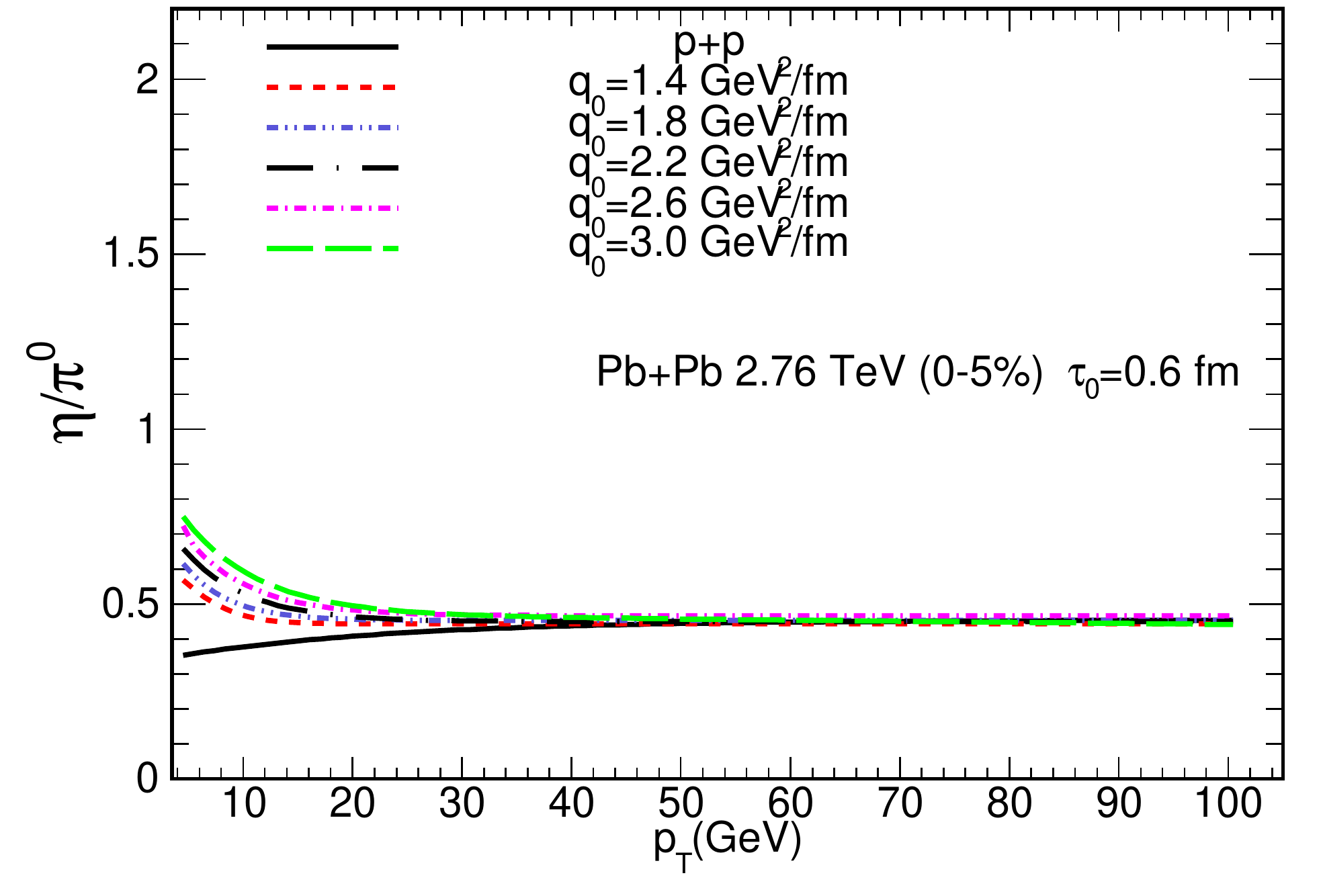}
\includegraphics[width=2.5in,height=2.in,angle=0]{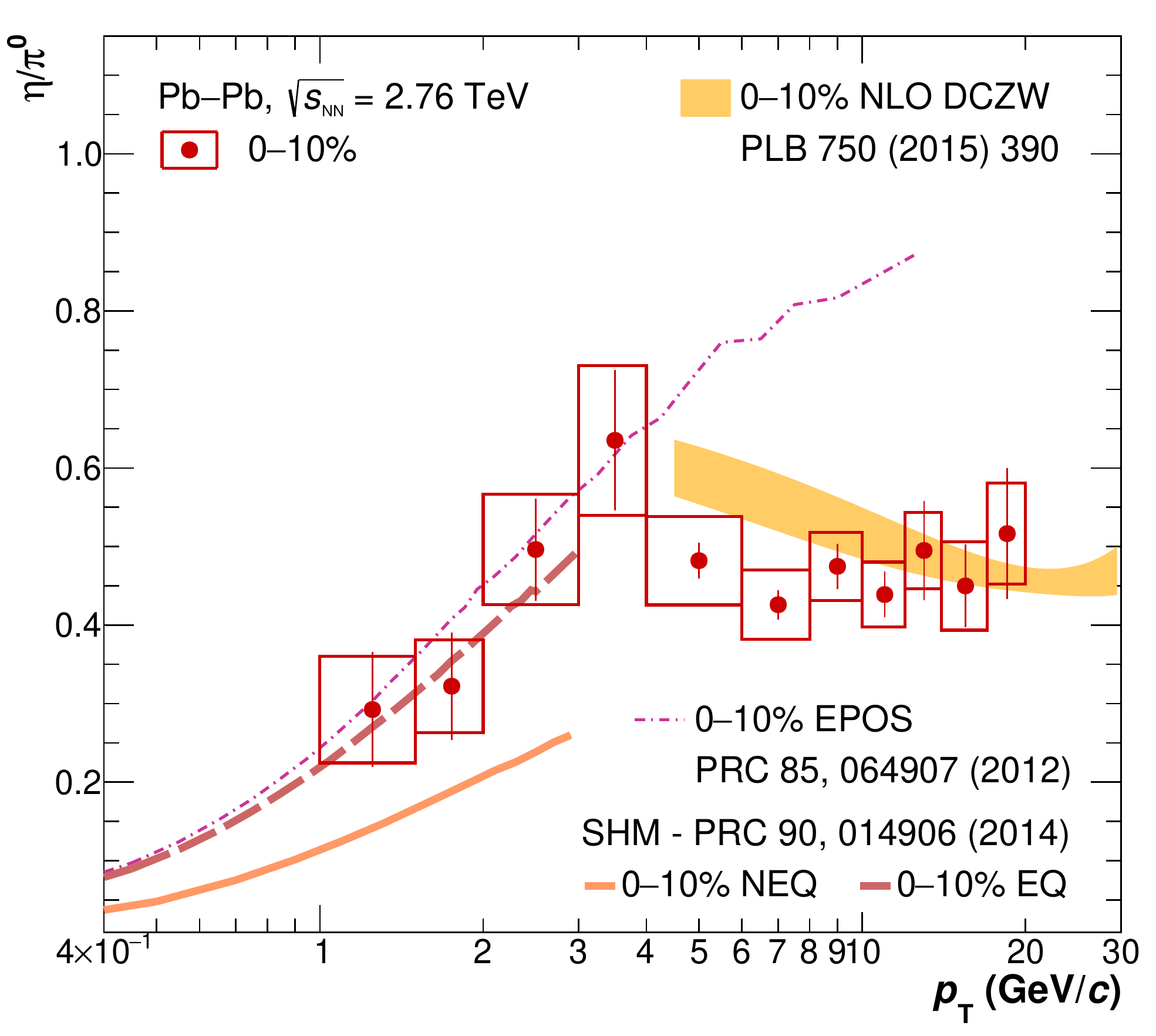}
\end{center}
\vspace*{-0.2cm}       
\caption{Left:   theoretical prediction from our model on the $\eta/\pi^0$ ratio at NLO in central Pb+Pb collisions at the LHC~\cite{Dai:2015dxa}.  Right: ALICE measurement of  $\eta/\pi^0$ ratio in the centrality class $0-10\%$ (circles)~\cite{Acharya:2018yhg} including a comparison with our model calculations (solid band); figure from ALICE~\cite{Acharya:2018yhg}.}
\label{fig:eta-2}       
\end{figure}

Recently we have investigated  productions of large momentum $\eta$ in p+p and A+A collisions at NLO~\cite{Dai:2015dxa}. In Fig.~\ref{fig:eta-1} (top panel) we show the contribution fractions from quarks and gluons to $\pi^{0}$ and $\eta$ yields in p+p at NLO at the RHIC. It is interesting to see that in p+p collisions quark (gluon) contribution fraction to $\pi^{0}$ is almost the same as quark contribution fraction to $\eta$, which means that the variation of quark (or gluon) fractions in the final-state partonic scattering will not change the yield ratio $\eta/\pi^0$.  Numerical simulations in the bottom panel of Fig.~\ref{fig:eta-1} (top panel) confirm this, which demonstrate that the ratio $\eta/\pi^0$ under the assumption with only gluons existing in the final-state partonic scattering, is very close to the one under the assumption with only quarks survives.

\begin{figure*}
\centering
\includegraphics[width=2.4in,height=2.2in,angle=0]{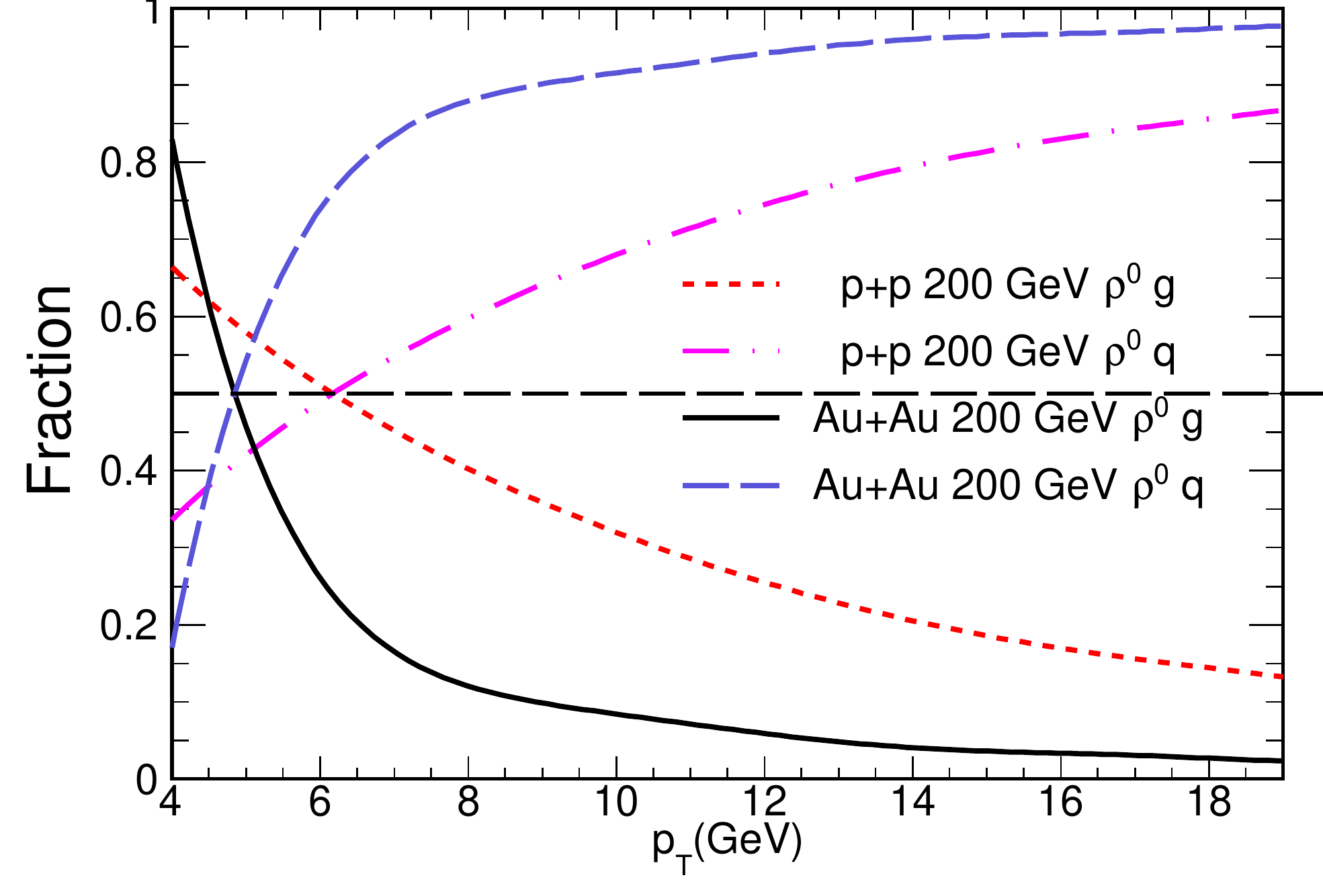}
\vspace*{1.8cm}       
\includegraphics[width=2.4in,height=2.2in,angle=0]{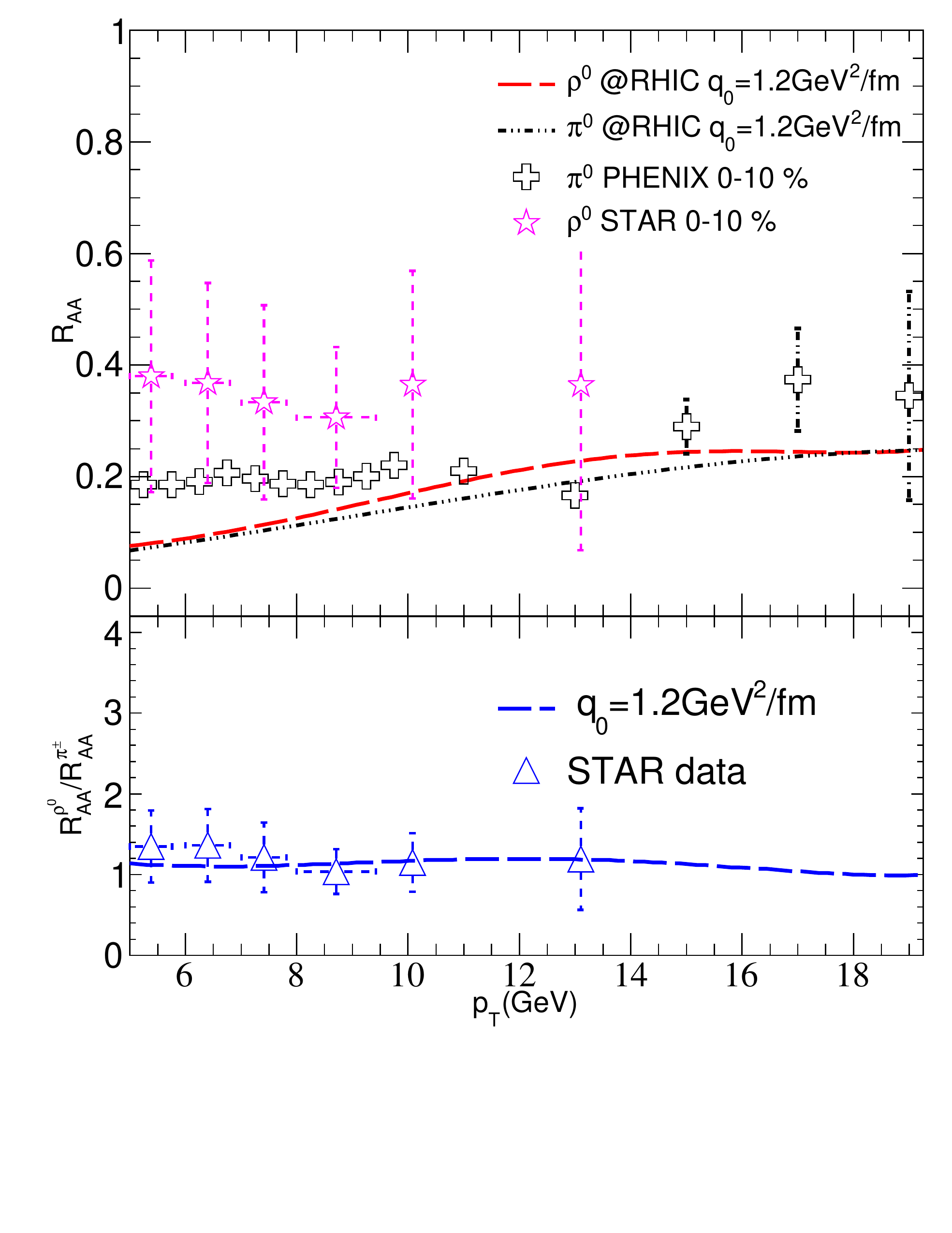}
\vspace*{-1.8cm}
\caption{Left: contribution fractions of $\rho^{0}$  yield from quark and gluon fragmentations at NLO in p+p and Au+Au at $\sqrt{s_{NN}}=200$~GeV~\cite{Dai:2017tuy}.
Right:  (top) nuclear modification factors of $\rho^{0}$ with comparison against data from STAR~\cite{Agakishiev:2011dc} and PHENIX~\cite{Adler:2003qi};  (bottom) double ratio  of $R_{\rm AA}^{\rho^0}/R_{\rm AA}^{\pi^\pm}$ at NLO, with comparing against STAR data.}
\label{fig:rho}       
\end{figure*}

Therefore even though in A+A collisions the quark fraction contribution will increase  because gluon should lose more energy than quark due to its large color charge, we expect this vairation of parton chemistry should not alter the ratio $\eta/\pi^0$, which is shown in Fig.~\ref{fig:eta-2} (left panel), where theoretical predictions on the yield ratio of $\eta/\pi^0$ in Pb+Pb at the LHC with different choices of jet transport coefficient $\hat{q}_0$ are presented. We find that  $\eta/\pi^0$ in Pb+Pb overlaps that in p+p in a wide range of $p_T$.  In 2018 ALICE published their measurement of neutral mesons in Pb+Pb at the LHC~\cite{Acharya:2018yhg}, and they compared the data on $\eta/\pi^0$ with our model predictions as illustrated in the right panel of Fig.~\ref{fig:eta-2}, where a decent agreement between our model calculation and the data is observed.

\begin{figure*}
\centering
\includegraphics[width=0.475\textwidth]{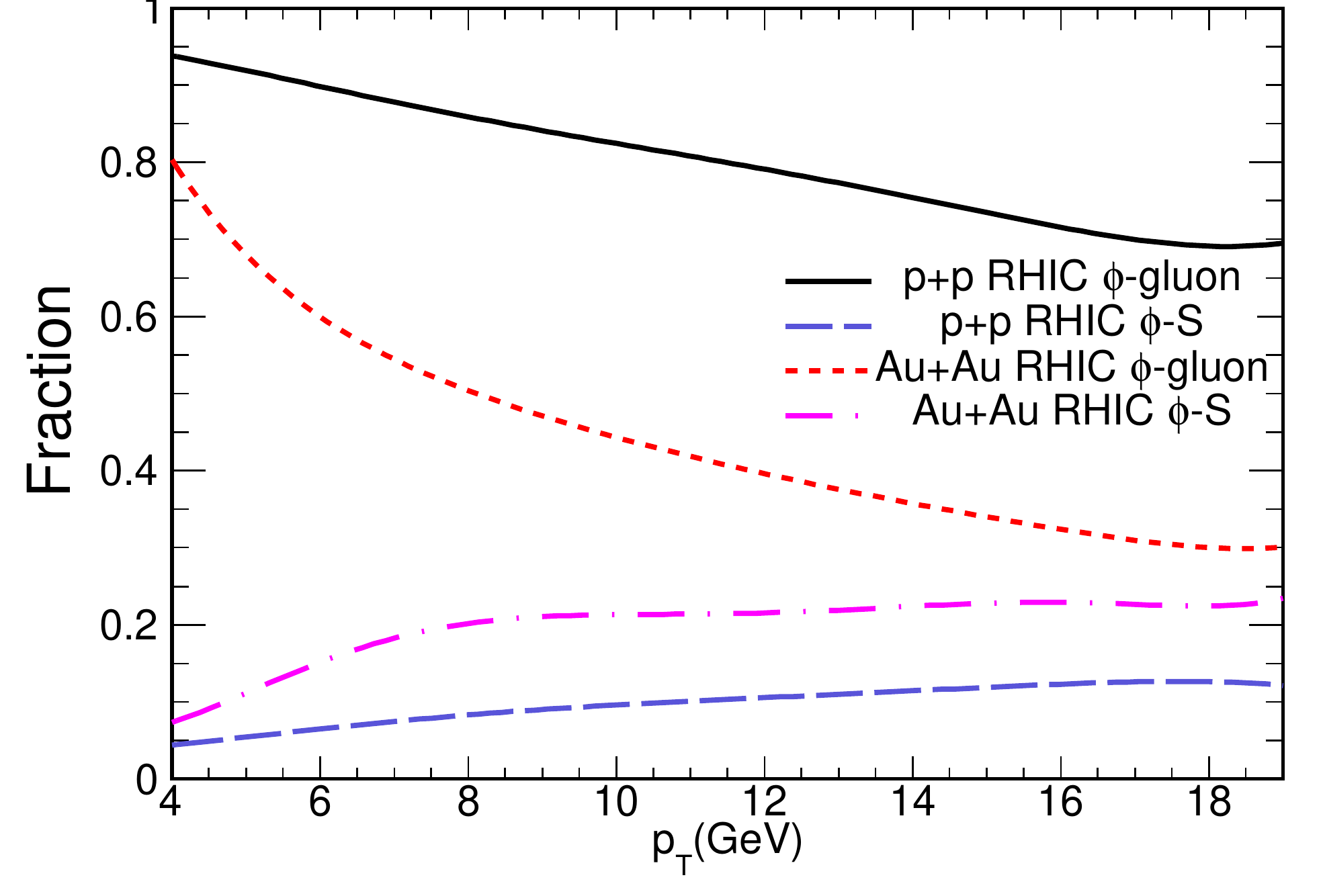}
\includegraphics[width=0.475\textwidth]{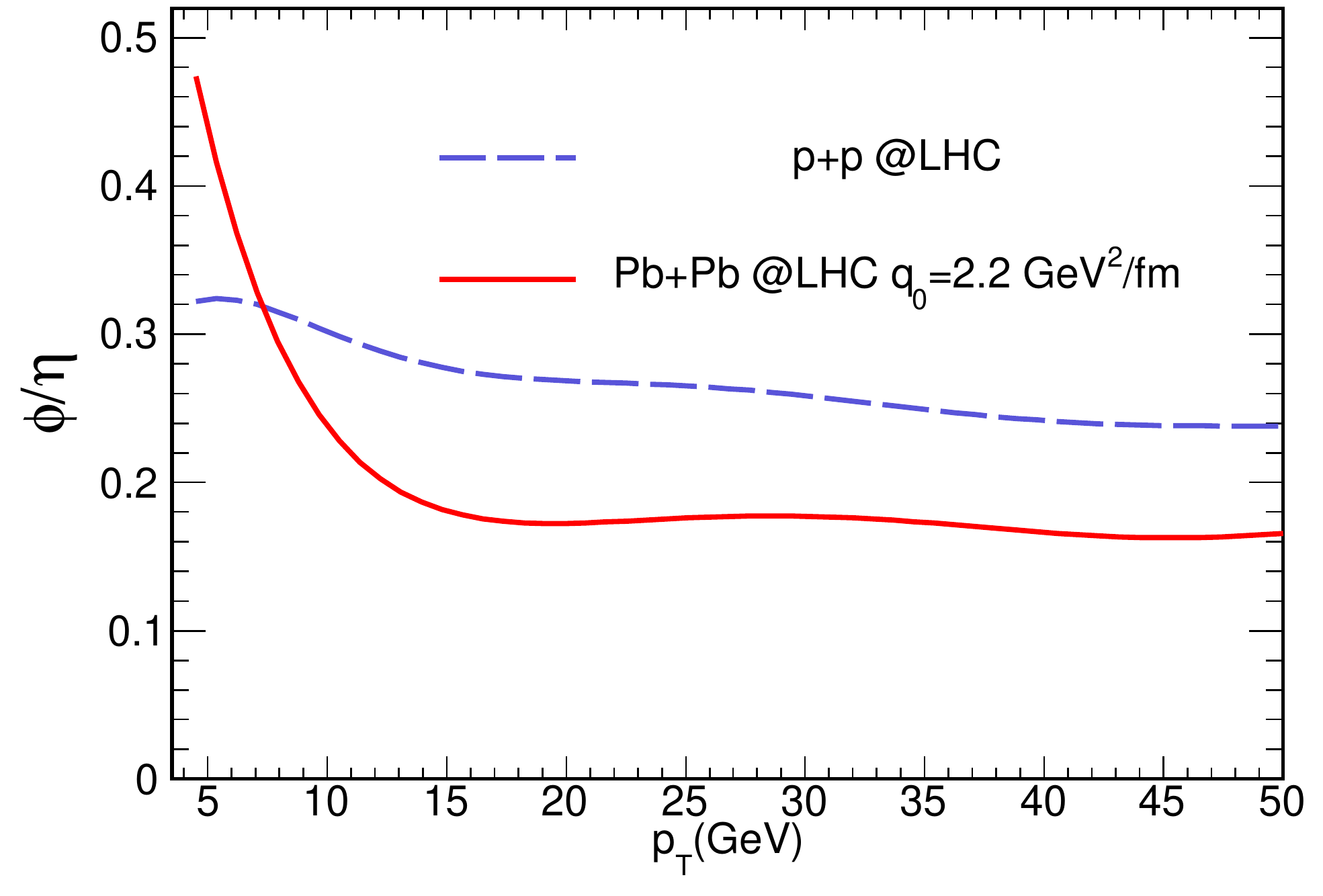}
\vspace*{0.1cm}       
\caption{Left: gluon and quark contribution fractions to the total $\phi$ yields in p+p and in
Au+Au at the RHIC~\cite{Dai:2017piq}. Right: the ratio of $\phi/\eta$ as a function of $p_T$
in p+p and Pb+Pb collisions at the LHC~\cite{Dai:2017piq}.}
\label{fig:phi}       
\end{figure*}

To take advantage of vacuum parton FFs of $\rho^0$ and $\phi$ from a broken SU(3)
model~\cite{Indumathi:2011vn,Saveetha:2013jda}, we also studied $\rho^0$ and $\phi$ mesons at large $p_T$ in A+A collisions at NLO with the same formalism~\cite{Dai:2017tuy, Dai:2017piq}.
The left panel of Fig.~\ref{fig:rho} shows that both in p+p and in A+A collisions with $p_T$ going up quark contribution fraction to $\rho^0$ increase, whereas gluon contribution fractions decrease; and the jet quenching effect will further reduce relative contribution from gluons.
In the right panel of  Fig.~\ref{fig:rho} we compare our calculations of nuclear modification factor for  $\rho^{0}$ meson at large $p_{\rm T}$ as well as the double ratio of $R_{\rm AA}^{\rho^0}/R_{\rm AA}^{\pi^\pm}$ at NLO with the experiment data. We can see the theory gives good descriptions on experiment data, especially for the double ratio $R_{\rm AA}^{\rho^0}/R_{\rm AA}^{\pi^\pm}$.

\begin{figure}[!h]
\begin{center}
\includegraphics[width=0.75\textwidth]{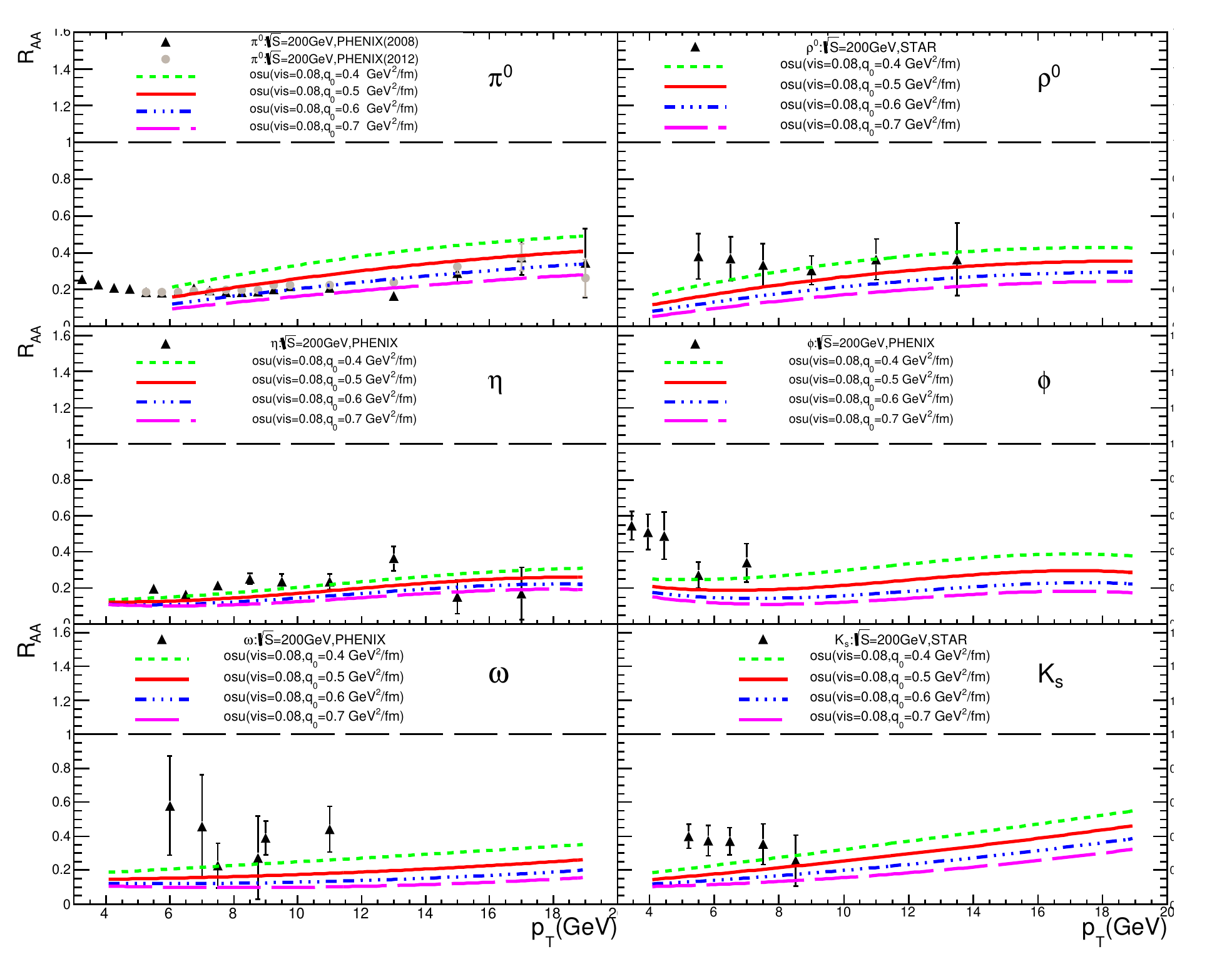}
\caption{Comparison of theoretical simulations of $R_{AA}$  at $\hat{q}_0=0.4 - 0.7$
~\cite{Ma:2018} with the RHIC experimental measurements of $\pi^0$~\cite{Adare:2008qa,Agakishiev:2011dc}, $\rho^0$~\cite{Agakishiev:2011dc}, $\eta$~\cite{Adare:2010dc}, $\phi$~\cite{Adare:2010pt},$\omega$~\cite{Adare:2011ht},$K^0_s$~\cite{Agakishiev:2011dc}.
}
\label{fig:fit-1}
\end{center}
\end{figure}

We plot the numerical results for $\phi$ meson in HIC in Fig.~\ref{fig:phi}. We see that contrary to the productions of $\pi$, $\eta$ and $\rho^0$, gluon contribution fraction to $\phi$ meson is dominant over
a wide $p_T$ region, and strange quark gives about $\sim 20$\% ($\sim 10$\%) contribution of $\phi$ meson yield in p+p (in Au+Au) though $\phi$ meson contains strange (and anti-strange) valence quark.  Because  $\pi$ meson (as well as $\eta$ and $\rho^0$) yield is dominant by quark fragmentation at very high $p_T$ in p+p, and the dominance of quark contribution is further enhanced by parton energy loss effect in A+A collisions, the ratio of $\phi/\eta$ in Pb+Pb shows a different behavior as the one in p+p in the $p_T$ region from
$15$~GeV to $50$~GeV (see right panel of Fig.~\ref{fig:phi}).

\begin{figure}[!h]
\begin{center}
\resizebox{0.475\textwidth}{!}{%
\includegraphics{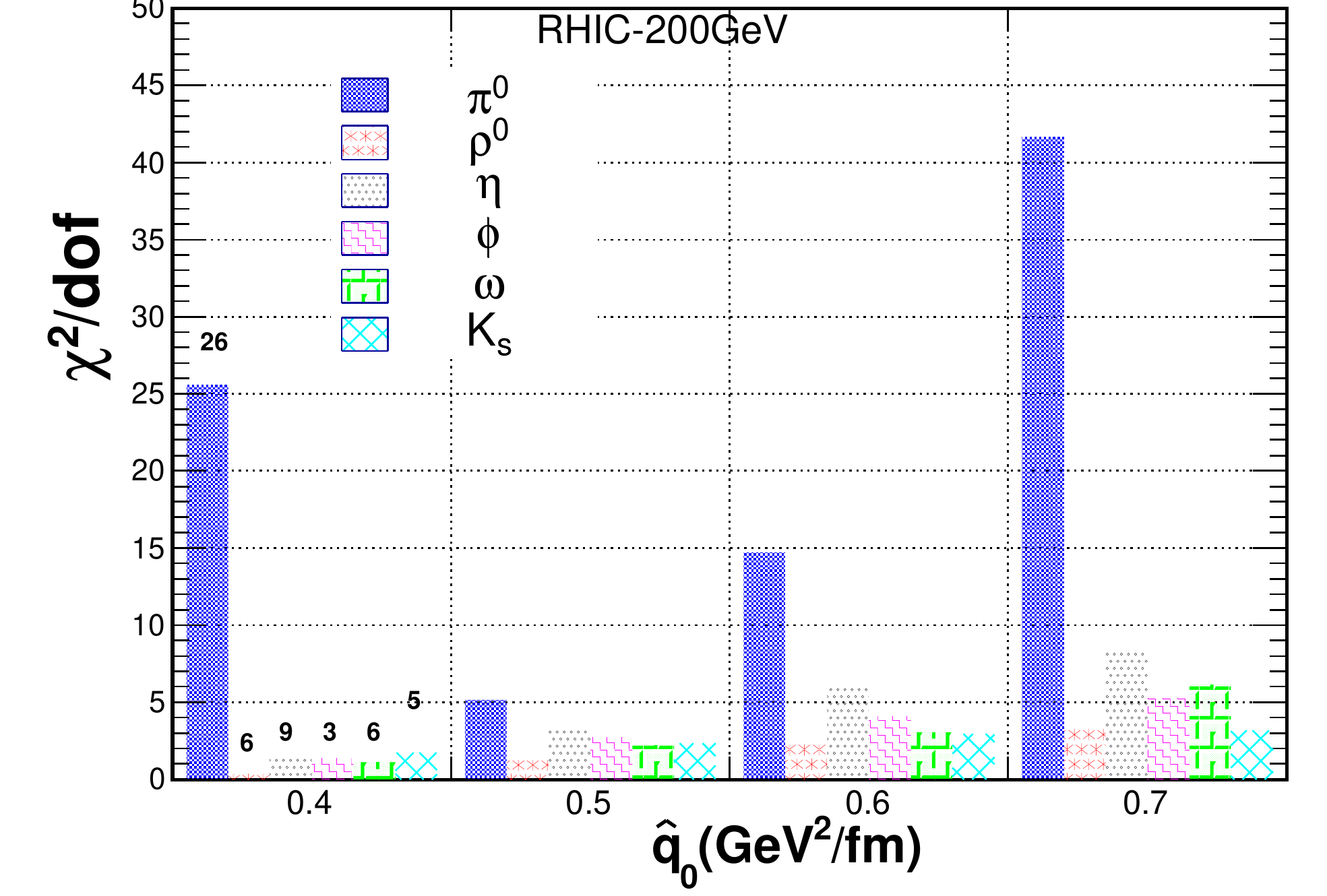}
}
\resizebox{0.475\textwidth}{!}{%
\includegraphics{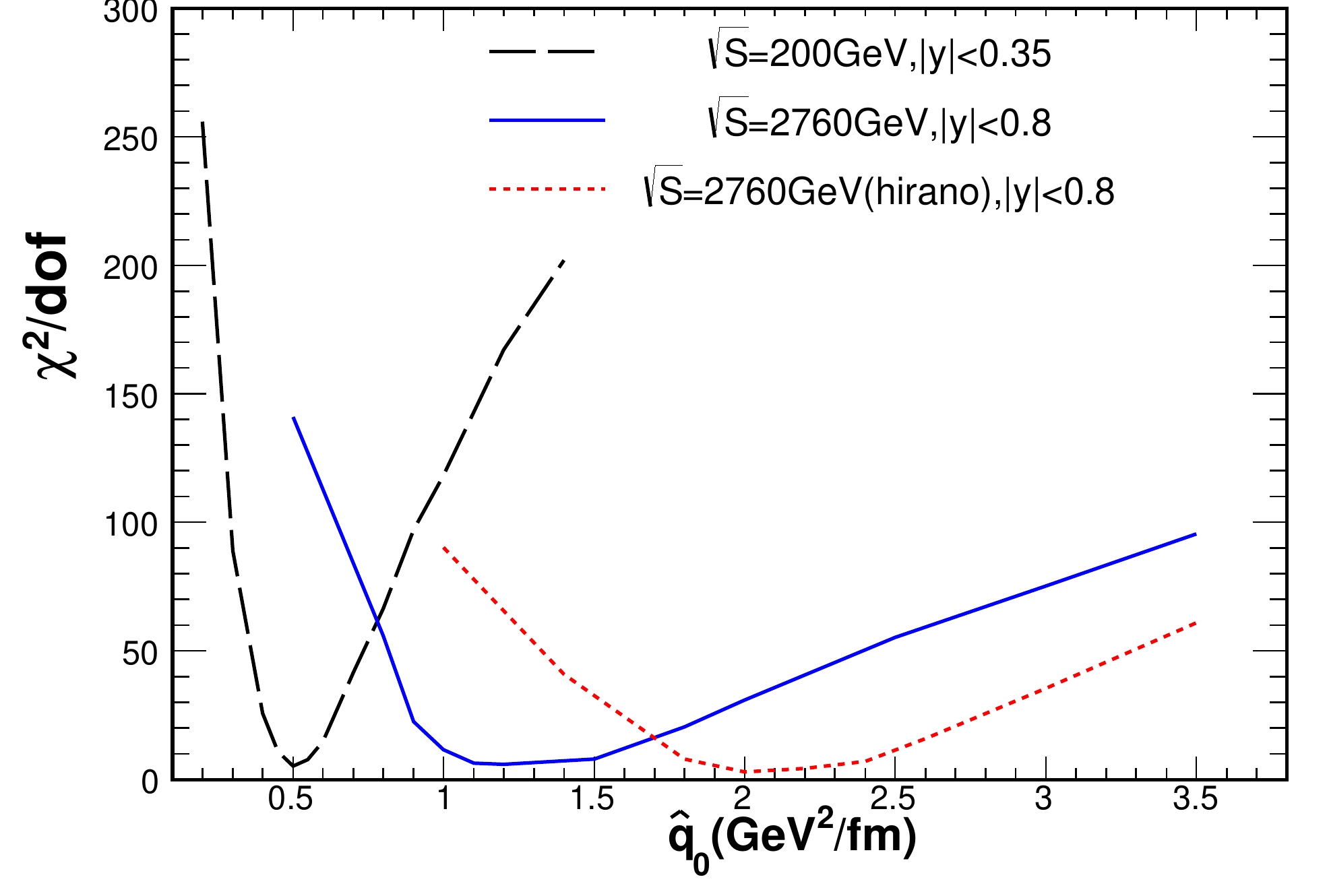}
}
\hspace*{-0.1in}
\caption{Left: the $\rm \chi^2/dof$ between the theoretical simulations of $\pi^0$, $\rho^0$, $\eta$, $\phi$, $\omega$, $K^0_{s}$'s $R_{\rm AA}$ at different $\hat q_0$ and the current public experimental data at the RHIC $200$~GeV~\cite{Ma:2018}. Right: the global $\rm \chi^2/dof$ of $R_{AA}$ with these 6 kinds of identified hadrons at the RHIC $200$~GeV and LHC $2.76$~TeV (with HIRANO hydro-description~\cite{Hirano:2002ds} and OSU iEBE-VISHNU hydro model~\cite{Shen:2014vra})
from a systematic study~\cite{Ma:2018}.
}
\label{fig:fit-2}
\end{center}
\end{figure}

Very recently we extend these studies of identified meson to the leading $\omega$ and $K^0_s$ meson productions in HIC at NLO. We note in the above discussion on productions of $\eta$, $\rho^0$ and $\phi$ the space-time evolution of the QGP is simulated with an ideal hydrodynamical model in Ref.~\cite{Hirano:2002ds}. With an event-by-event viscous hydrodynamical model iEBE-VISHNU~\cite{Shen:2014vra} we calculate the productions of six types of identified hadrons
($\pi^0$, $\rho^0$, $\eta$, $\phi$, $\omega$, and $K^0_s$) in A+A collisions at NLO by considering jet quenching effect with high-twist approach~\cite{Ma:2018}, and some selected results are shown in
Fig.~\ref{fig:fit-1} and Fig.~\ref{fig:fit-2}.

In Fig.~\ref{fig:fit-1}, We perform a systematic calculation of  nuclear modification factors $\rm R_{AA}$ of six identified mesons ($\pi^0$, $\rho^0$, $\eta$, $\phi$, $\omega$, and $K^0_s$)  in Au+Au collisions and make comparison with all available experimental data at RHIC.

In order to extract the best value of  the jet transport parameter  $\hat{q}_0$ at the initial time of the QGP formation, we perform a global $\chi^2$ fitting by comparing the theoretical results on productions of six types of mesons at different values of $\hat{q}_0$ with the available experimental data as:
\begin{eqnarray}
\chi^2({a_j}) = \sum_i \frac{[D_i-T_i({a_j})]^2}{\sigma_i^2}
\label{eq:chi2}
\end{eqnarray}

\hspace{-0.2in}

Fig.~\ref{fig:fit-2} (left panel) gives the global $\chi^2/\rm dof$ for large momentum yields of $6$ mesons with various choices of jet transport coefficient $\rm \hat{q}_0 =0.4 - 0.7 GeV^2/s$ at the RHIC $\sqrt{s}_{\rm NN}=200$~GeV.
Fig.~\ref{fig:fit-2} (right panel)  illustrate that the minima of $\rm \chi^2/dof$ of $\pi^0$ at the RHIC and the LHC respectively give the best fitting of $\rm \hat{q}_0=0.5~GeV^2/fm$ at the RHIC, and
$\rm \hat{q}_0=1.1-1.4~GeV^2/fm$ at the LHC with OSU iEBE-VISHNU hydro model~\cite{Shen:2014vra}. It is noted that the extraction of jet transport coefficient is sensitive to the QGP fireball evolution, and the corresponding calculation with the hydro model by Hirano et al~\cite{Hirano:2002ds} may give larger $\hat{q}_0$.


\section{Full Jet Observables in HIC}
\label{Full Jet Observables}

\begin{figure}
  \centering
\vspace*{-0.4in}
\includegraphics[width=3.5in,height=2.5in,angle=0]{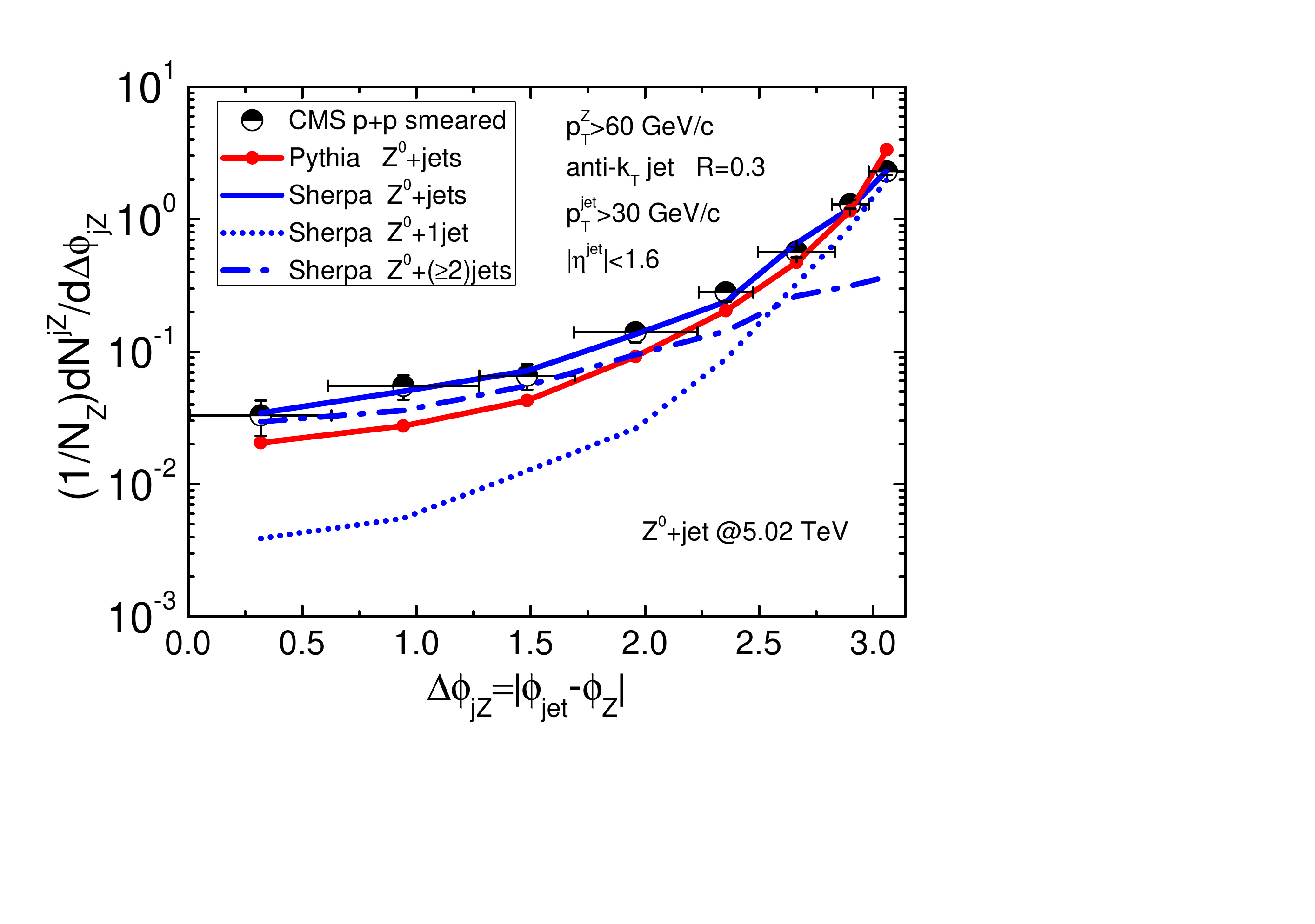}
\vspace*{-0.6in}
  \caption{The azimuthal angle correlations $\Delta\phi_{jZ}$ in p+p collisions at 5.02 TeV
  with the NLO+PS simulations by Sherpa, and LO+PS simulations by Pythia~\cite{Zhang:2018urd}. Experimental data are from CMS measurement ~\cite{Sirunyan:2017jic}.}
  \label{fig:Z-1}
\end{figure}

\subsection{$Z^0$/$W^{\pm}$+jet}
\label{Z/W+jet}
The process of gauge boson associated jet production has been regarded as a ``golden channel'' to study jet quenching because gauge boson doesn't participant in the strong interaction and its energy could then be used to calibrate the initial energy of tagged jet before propagating in the QGP~\cite{Neufeld:2010fj, Kang:2017xnc}.  Recently CMS collaboration has measured $Z^0$+jet azimuthal angle correlation $\Delta \phi_{\rm jZ}= |\phi_{\rm jet}-\phi_{Z}|$  in p+p and Pb+Pb
collisions~\cite{Sirunyan:2017jic}. We notice that the fixed-order calculation on the azimuthal angle distribution may fail near the region $\Delta \phi_{\rm jZ} \sim \pi$ due to soft/collinear divergence; whereas higher-order correction beyond leading-order (LO) contribution is important in small and intermediate $\Delta \phi_{\rm jZ}$ region.

\begin{figure}[tpb]
 \begin{center} \centering
   \includegraphics[scale=0.2]{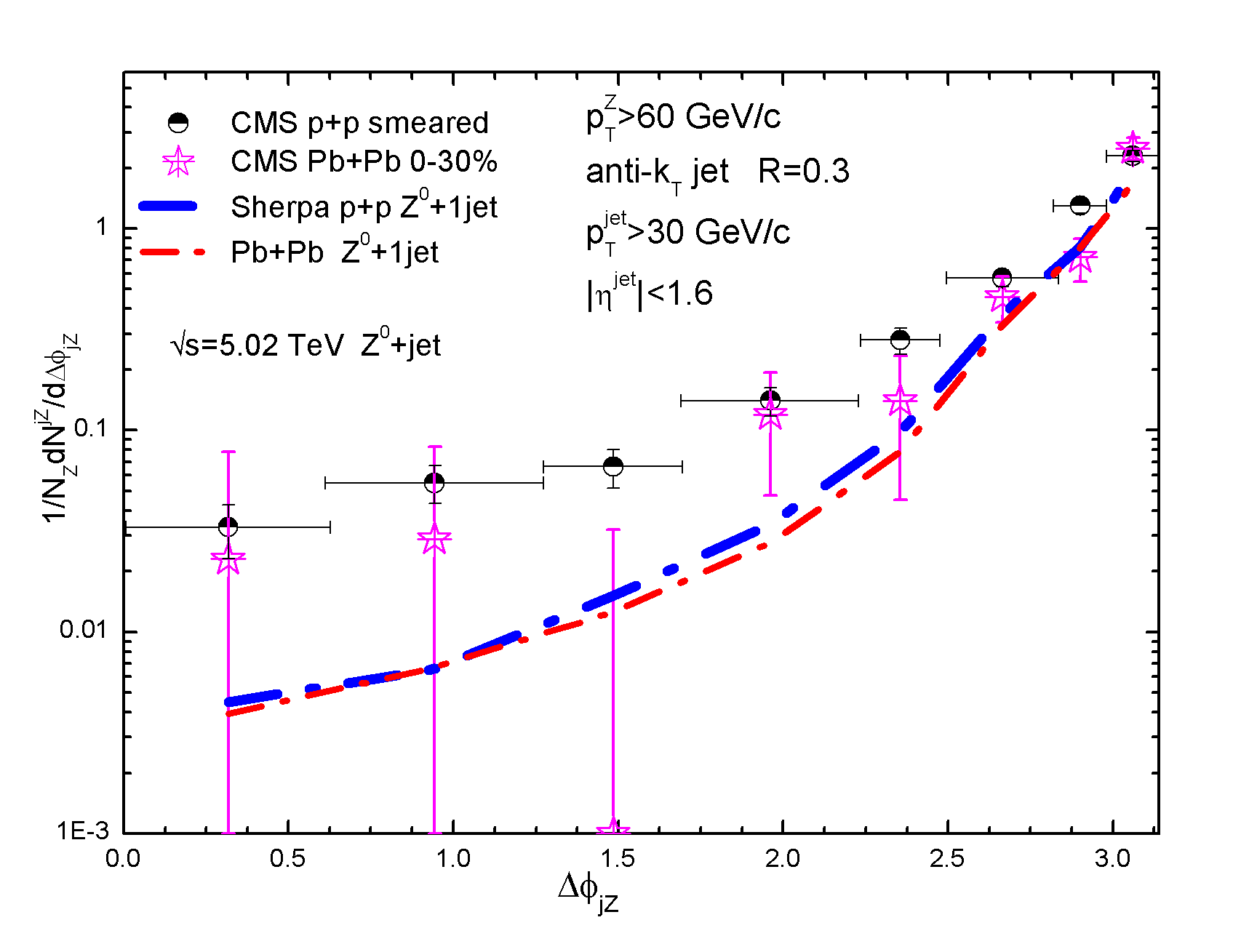}
   \includegraphics[scale=0.2]{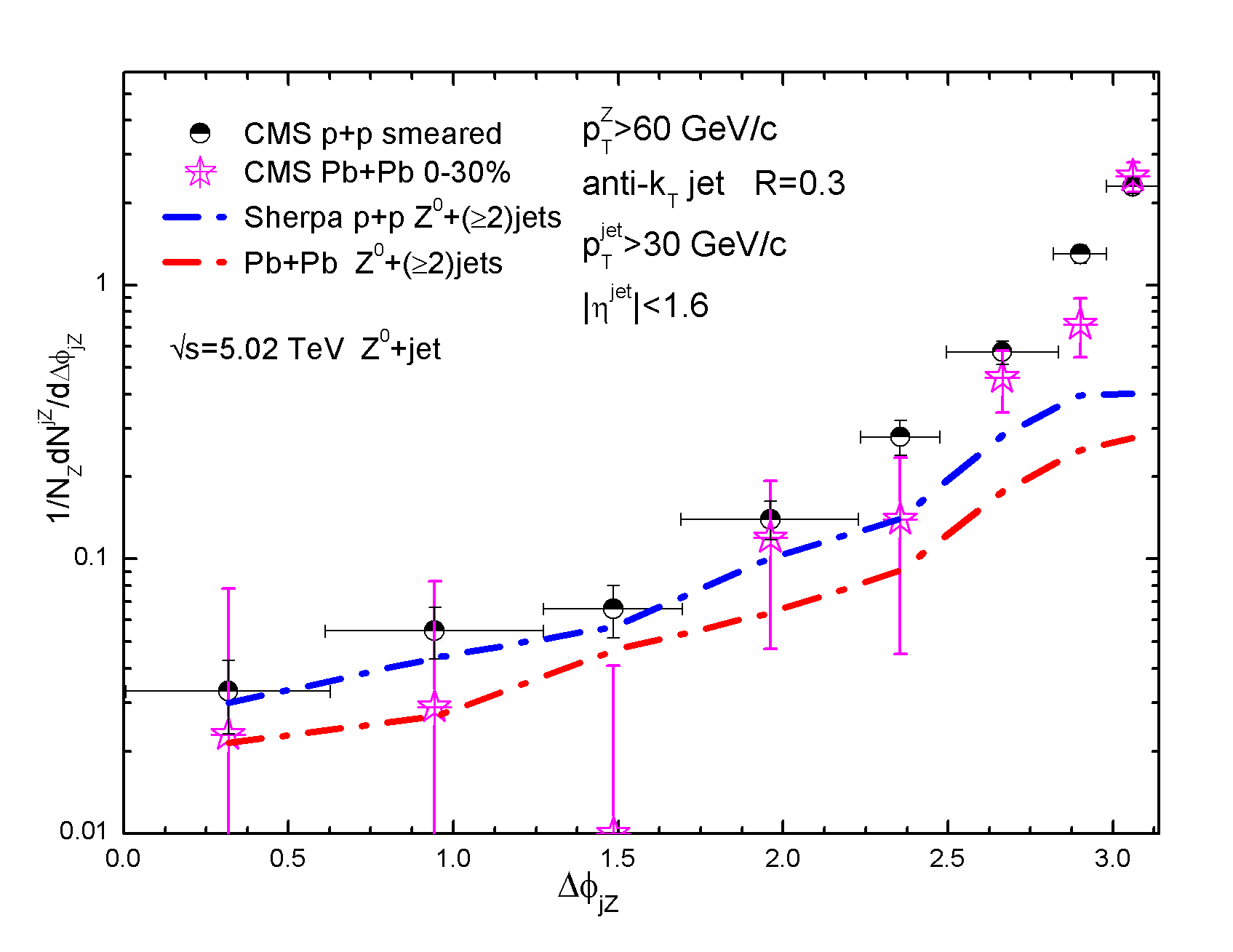}
 \end{center}
   \vspace{-10pt}
  \caption{Azimuthal angle distributions of $Z^0$ tagged jet production when considering the processes of  $Z^0$ plus only one jet (left), and $Z^0$ plus more than one jet (right) both in p+p and central Pb+Pb collisions at $\sqrt{s_{NN}}=5.02$~TeV.   }
  \label{fig:Z-2}
\end{figure}

To study the angle correlation of $Z^0$+jet we developed a formalism that combines the NLO matrix elements with resummation of a matched parton shower (PS) for the initial $Z^0$ associated jet production, and includes jet quenching effect in the QGP~\cite{Zhang:2018urd}.
In Fig.~\ref{fig:Z-1} we plot the comparison  the theoretical results from
a NLO+PS Monte Carlo event generator, Sherpa~\cite{Gleisberg:2008ta}, on azimuthal angle correlation of jet production associated with $Z^0$ in p+p collisions with experimental data, and a very nice agreement between theory and data is observed, whereas Pythia with LO+PS undershoots the data at small $\Delta \phi_{\rm jZ}$ where higher order processes $Z^0+(\geq 2)$ jets give significant contribution.

\begin{figure}
  \centering
   \includegraphics[scale=0.22]{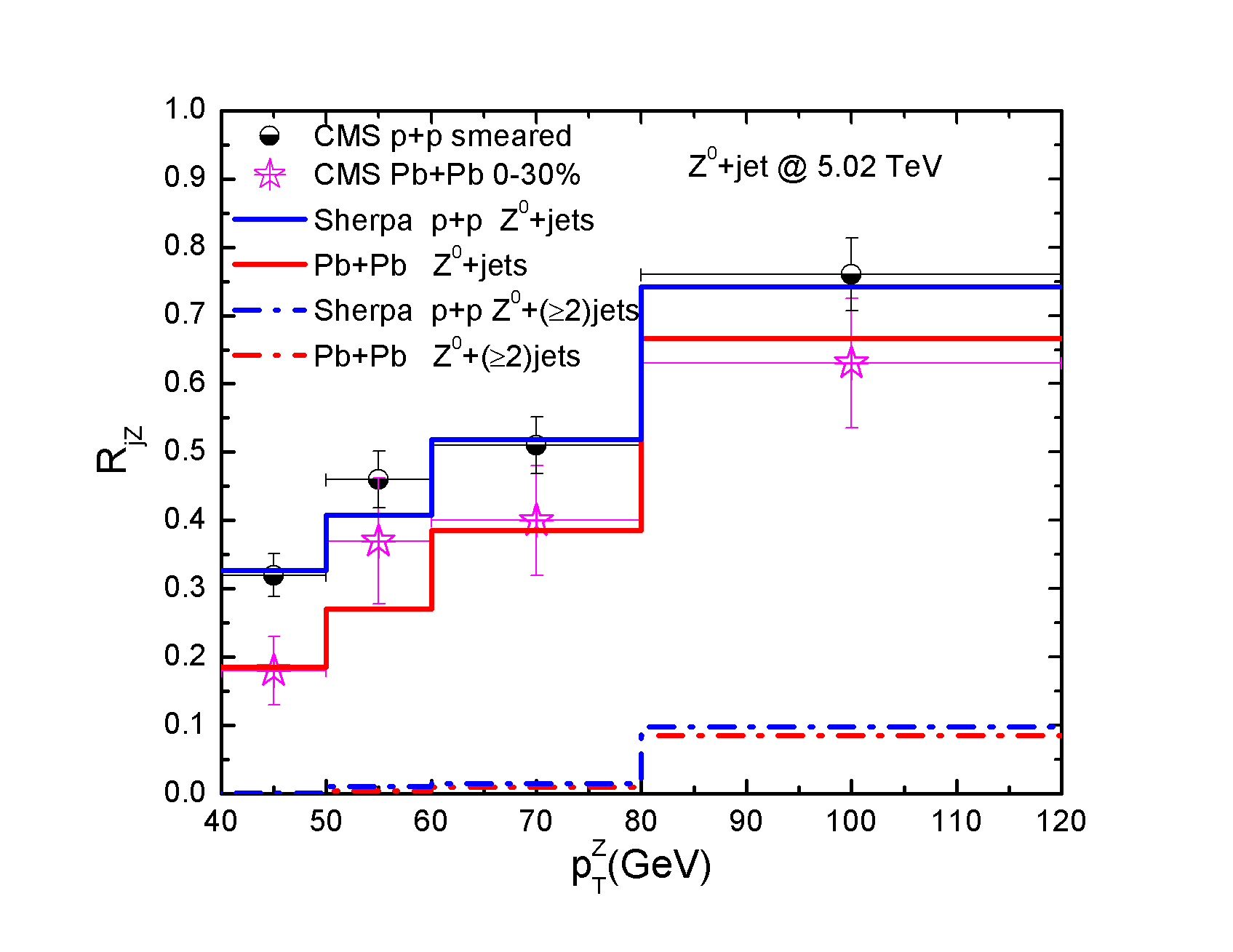}
    \vspace{-10pt}
  \caption{Distributions of  averaged number of jet per Z boson $R_{jZ}$
  in  p+p and central Pb+Pb collisions at the LHC.
  }\label{fig:Z-3}
\end{figure}

We utilize the LBT model to simulate the propagation of energetic partons and its energy loss in the QGP which includes both elastic and inelastic processes of parton scattering in the QCD medium~\cite{He:2015pra}. In Fig.~\ref{fig:Z-2}, we present medium modifications on Z+jet azimuthal angle correlations of $Z^0+ 1$ jet (left panel) and  $Z^0+(\geq 2)$ jets (right panel) in Pb+Pb collisions  at the LHC. It is found that the jet quenching effect will give rather modest medium modification of angle correlations for $Z^0+ 1$ jet processes, but considerable suppression for  $Z^0+(\geq 2)$ jets processes because jets in these processes have small initial energies and are more ready to fall below the threshold cut $p_T \geq 30$~GeV after losing energies in the QGP. Fig.~\ref{fig:Z-3} gives average number of jet per Z boson $R_{jZ}$ as a function of $p_T^{Z}$ in p+p and Pb+Pb at the LHC. It shows that $R_{jZ}$ is reduced in Pb+Pb relative to that in p+p due to parton energy loss.

\begin{figure}[tpb]
 \begin{center} \centering
   \includegraphics[width=0.8\textwidth]{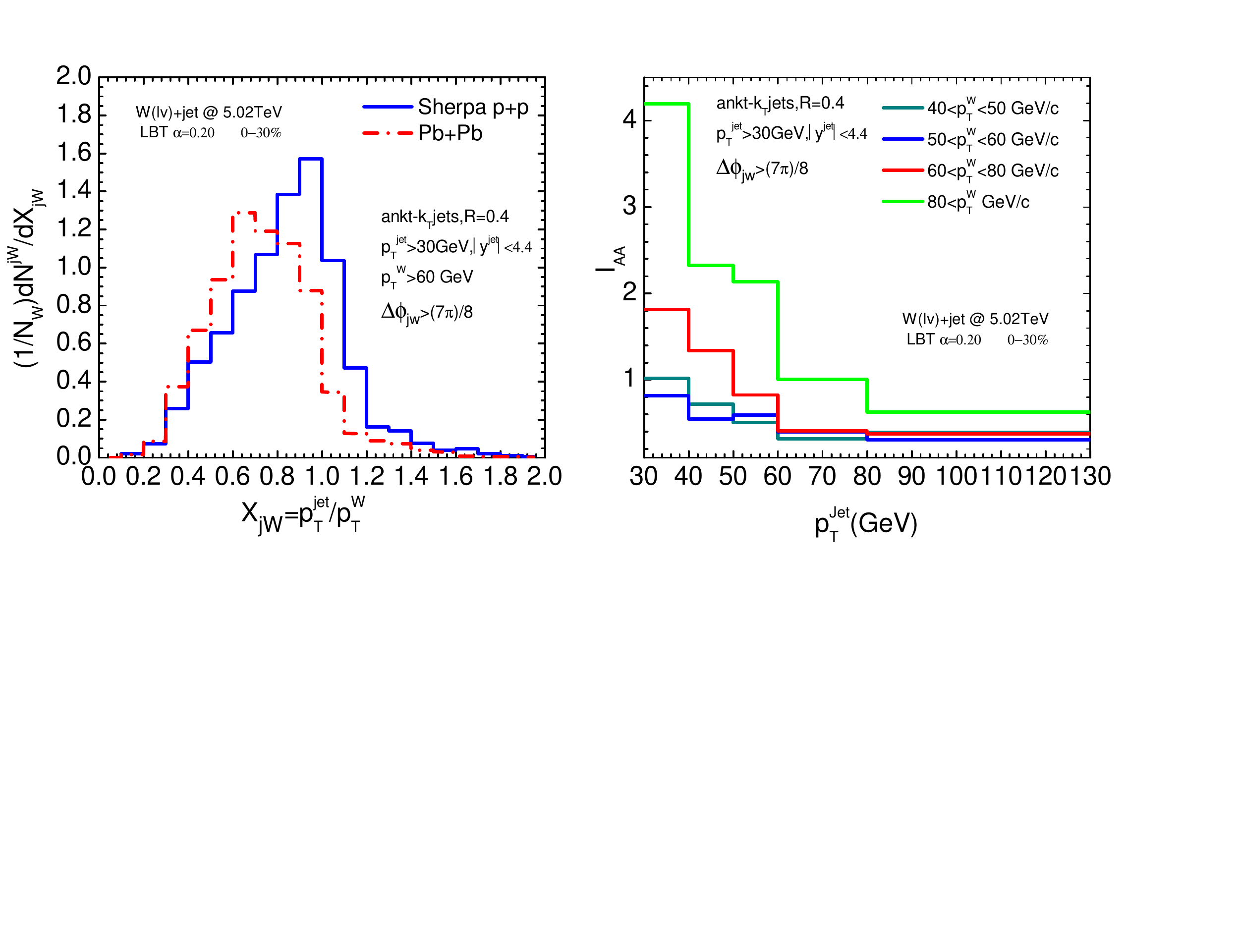}
 \end{center}
   \vspace{-120pt}
  \caption{Left: the transverse momentum imbalance of W+jet in p+p and Pb+Pb at the LHC.
   Right:  $I_{\rm AA}$ of  W+jet as a function of $p_T^{\rm jet}$ at the LHC.
  }\label{fig:W}
\end{figure}

With the same formalism of NLO+PS+Eloss we just finish a study of W+jet in Pb+Pb at the LHC, and the numerical results are show in Fig.~\ref{fig:W}~\cite{Zhang:2018}. Generally speaking the results of W+jet in HIC are similar to those of Z+jet in HIC. The left panel of Fig.~\ref{fig:W} gives momentum imbalance $x_{jW}=p^{\rm jet}_T/p^W_T$ distribution of W+jet in p+p and Pb+Pb collisions at the LHC, and shows that the jet quenching effect will shift momentum imbalance spectrum to the left.
The right panel of Fig.~\ref{fig:W} presents the ratio~\cite{Neufeld:2010fj},

$$ I_{\rm AA}=\frac{1}{\langle N_{\rm bin}\rangle} \left.  \frac{d\sigma^{AA}}{dp^W_T dp^{\rm jet}_T}
\right/  \frac{d\sigma^{pp}}{dp^W_T dp^{\rm jet}_T} $$

as a function of  jet transverse momentum  $p_T^{\rm jet}$ at different $p^W_T$ regions. We clearly see an enhancement at small  $p_T^{\rm jet}$ and a suppression at large  $p_T^{\rm jet}$, as we observed already in Z+jet productions~\cite{Neufeld:2010fj}.

We have checked that in the above calculations the cold nuclear matter effects are rather small in the kinematic regions we are interested in~\cite{Ru:2014yma}.

\subsection{Heavy flavor dijet in HIC}
\label{Double b-jet}

Flavor dependence of parton energy loss is an active research topic of jet quenching, and the production of
 $b \bar{b}$ dijet in HIC will help understand the mass dependence of  parton energy loss.  The pair of $b \bar{b}$ jets or double b-tagged jets refer to processes with the final-state two jets which contains one bottom quark or anti-bottom quark. With a similar approach of computing Z/W+jet production in Pb+Pb, we employ NLO+PS event generator Sherpa to describe the initial hard processes of double b-tagged jets, and Langevin transport equations to consider the propagating of the heavy quark in the QGP, to make a first theoretical study of $b \bar{b}$ dijet in HIC ~\cite{Dai:2018mhw}.  This NLO+PS+Eloss model can give nice descriptions of data on inclusive jet, light flavor dijet and $b \bar{b}$ dijet both in p+p and Pb+Pb collisions  at the LHC~\cite{Sirunyan:2018jju}.

\begin{figure}[!t]
\begin{center}
\includegraphics[width=0.4\textwidth,angle=0]{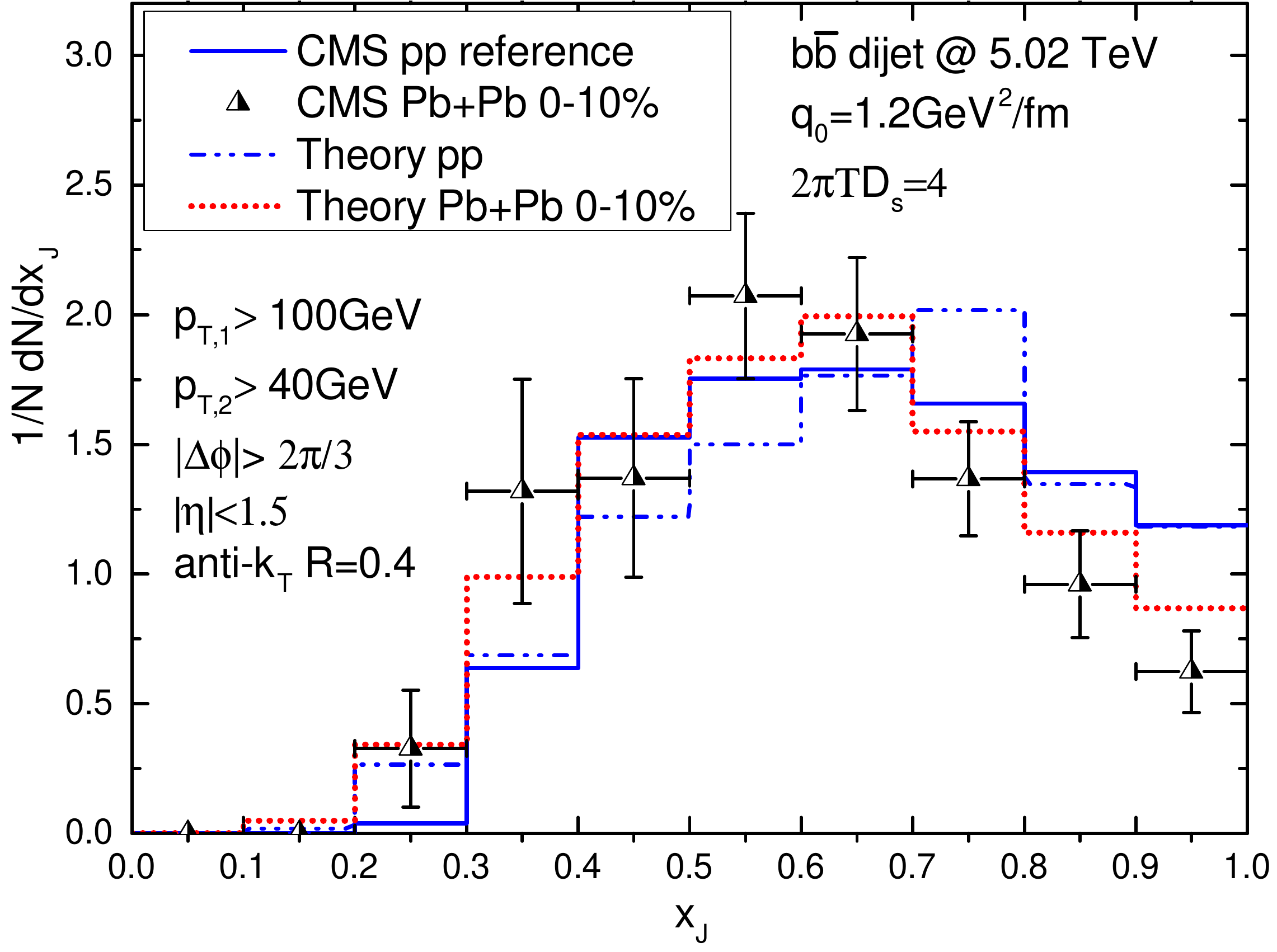}
\includegraphics[width=0.4\textwidth,angle=0]{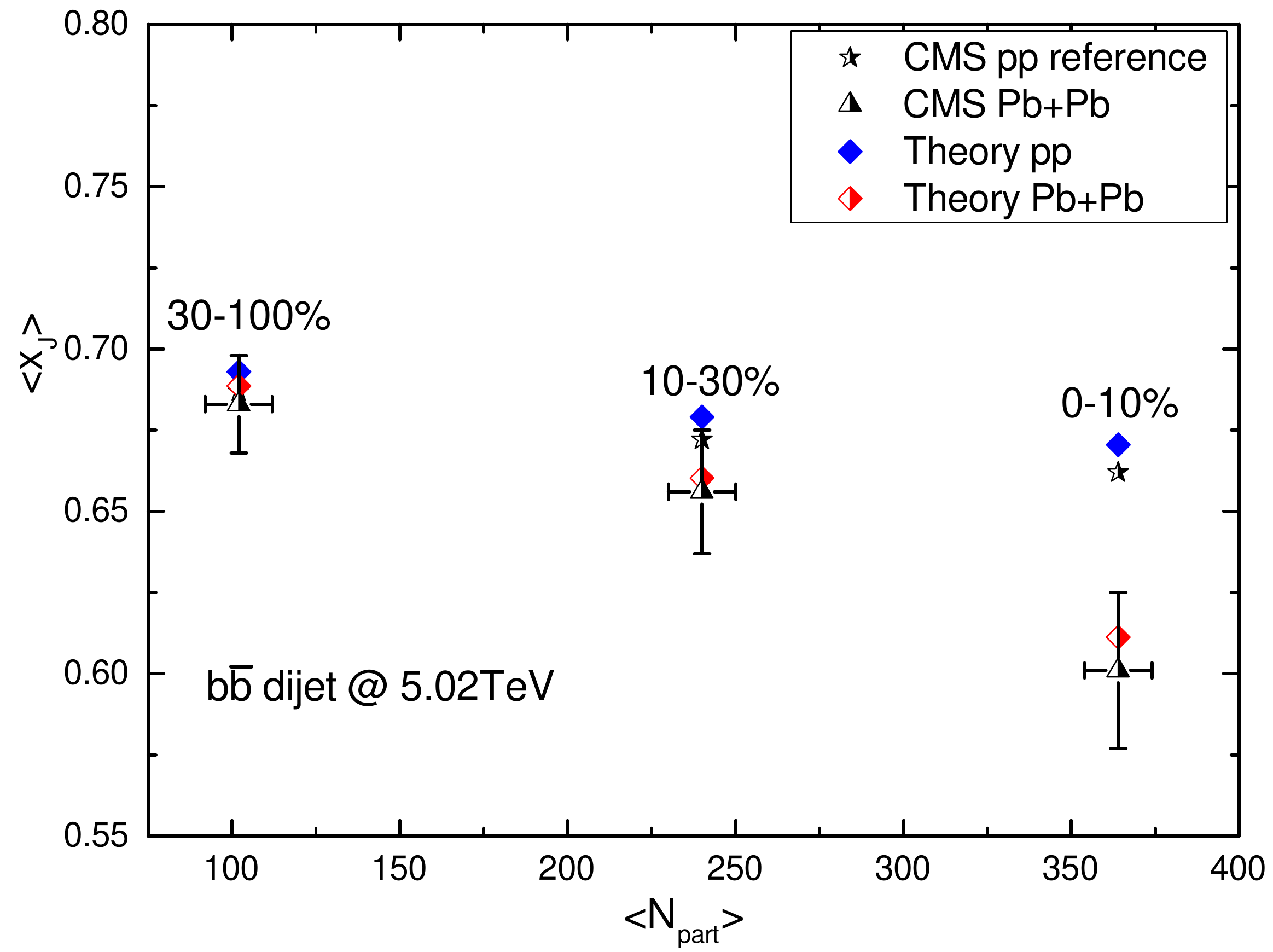}
\caption{
Left: normalized $x_J$ distribution of  heavy flavor dijet in p+p and central Pb+Pb collisions
compared with the smeared p+p baseline and experimental data in A+A collisions.
Right: averaged values of $x_J$ of $b \bar{b}$ dijet production at different centralities compared with experimental p+p references and Pb+Pb data. The data are taken from CMS measurement~\cite{Sirunyan:2018jju}.
}
\label{fig:bjet-1}
\end{center}
\end{figure}


In Fig.~\ref{fig:bjet-1}, we plot the distribution of momentum imbalance $x_J=p_{\rm T,2}/p_{\rm T,1}$,  the $p_T$ ratio of the sub-leading b-jet to leading b-jet for $b \bar{b}$ dijet by following the kinematic cuts adopted by CMS experiment, where the minimum transverse momenta of the leading and the sub-leading b-tagged jets are set to be $100$~GeV and $40$~GeV respectively. One can observe
that momentum imbalance distribution in Pb+Pb is shifted to the left, i.e.  to the smaller values relative to p+p collisions, and this shifting becomes larger in most central collisions, as demonstrated also in the right panel of   Fig.~\ref{fig:bjet-1}, which shows the depletion of averaged $x_J$ in Pb+Pb as compared to those in p+p increases with the number of number of participant.

In Fig.~\ref{fig:bjet-1}, we present the azimuthal angle correlation of a pair of b-jets,
requiring  the transverse momentum of b-jets $p_{\rm T} \geq 15$~GeV.  Here $\Delta \phi$ is the azimuthal angle difference between two b-quark tagged jets. It is found that
jet quenching effect may give a modest suppression to the angle distribution in the small $\Delta \phi$ region, and an enhancement in the large $\Delta \phi$ region, quite similar to what we observed
in Z+jet productions. We emphasize that azimuthal angle correlations of $b\bar{b}$ dijets are sensitive to the value of the kinematic cut on b-jet transverse momentum, and they may exhibit quite distinct behaviors with different choices of  $p_T$ cut  in p+p and A+A collisions~\cite{Dai:2018mhw}.
We notice that the production of $b\bar{b}$ dijets in HIC has also been studied very recently in Ref.~\cite{Kang:2018wrs}.

\begin{figure}[!t]
\begin{center}
\vspace*{-0.2in}
\hspace*{-.1in}
\includegraphics[width=2.2in,height=2.in,angle=0]{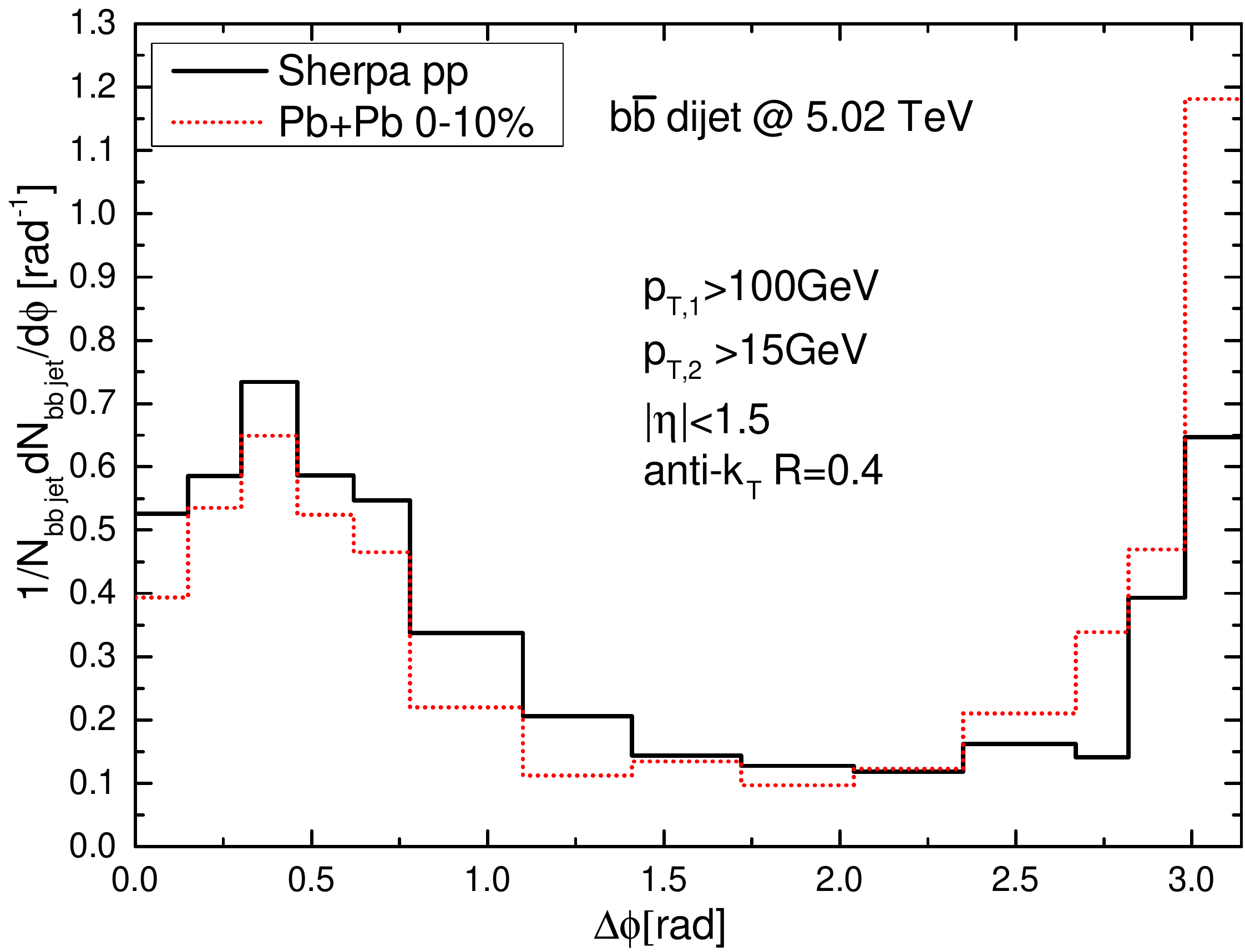}
\vspace*{-.2in}
\caption{
Azimuthal angle distributions of $b\bar{b}$ dijets in p+p and Pb+Pb collisions at the LHC with minimum cut of b jet momentum $p_{\rm T}=15$~GeV.
}
\label{fig:bjet-2}
\end{center}
\end{figure}

%
%
\vspace*{.3cm}
{\bf Acknowledgments:}  This research is supported by the NSFC of China with Project
Nos. 11435004, 11805167.

\end{document}